\documentclass[superscriptaddress,english,pre,showpacs,longbibliography]{revtex4-2}

\usepackage{titlesec}

\renewcommand{\thesection}{\arabic{section}} %
\renewcommand{\thesubsection}{\thesection.\arabic{subsection}} %
\renewcommand{\thesubsubsection}{\thesubsection.\arabic{subsubsection}} %

\titleformat{\section}
  {\normalfont\Large\bfseries}
  {\thesection.}
  {1em}
  {}

\titleformat{\subsection}
  {\normalfont\normalsize\bfseries}
  {\thesubsection.}
  {1em}
  {}

\titleformat{\subsubsection}
  {\normalfont\normalsize\bfseries}
  {\thesubsubsection.}
  {1em}
  {}

\usepackage{listings}
\usepackage{xcolor}

\definecolor{codebg}{rgb}{0.95,0.95,0.95} %
\definecolor{codecomment}{rgb}{0,0.6,0} %
\definecolor{codestring}{rgb}{0.58,0,0.82} %
\definecolor{codekeyword}{rgb}{0,0,1} %

\usepackage[colorlinks=true,urlcolor=blue,citecolor=blue,linkcolor=blue]{hyperref}
\usepackage{makecell}
\usepackage{enumitem} 
\usepackage{graphicx}
\usepackage{dcolumn}
\usepackage{bm}
\usepackage{amsmath,amssymb,amsthm,mathrsfs,amsfonts,dsfont}
\usepackage{balance}
\usepackage{color}
\usepackage{array}
\usepackage{tabularx}
\usepackage{longtable}
\usepackage{color}
\usepackage{float}
\usepackage{indentfirst}
\usepackage{txfonts}
\usepackage{booktabs} %
\usepackage{multirow}

\usepackage{tocvsec2}
\usepackage{placeins}

\usepackage{physics}
\usepackage{algorithm} 
\usepackage{algorithmic}
\usepackage{subfloat}
\usepackage{graphicx}

\usepackage[all]{hypcap}
\makeatletter
\let\saved@includegraphics\includegraphics
\AtBeginDocument{\let\includegraphics\saved@includegraphics}
\makeatother
\usepackage{braket}
\usepackage{tikz}
\usepackage{float}
\usepackage{amsmath,amssymb,amsthm,mathrsfs,amsfonts,dsfont}

\usepackage{xcolor}

\newcommand{\s}{\bm{s}}

\begin{document}
\title{BatchTNMC: Efficient sampling of two-dimensional spin glasses using tensor network Monte Carlo}

\author{
Tao Chen
}
\thanks{These authors contributed equally}
\affiliation{Hefei National Laboratory for Physical Sciences at the Microscale and Department of Modern Physics, University of Science and Technology of China, Hefei 230026, China}
\affiliation{Hefei National Laboratory, University of Science and Technology of China, Hefei 230088, China}

\author{
Jingtong Zhang
}
\thanks{These authors contributed equally}
\affiliation{Institute of Theoretical Physics, Chinese Academy of Sciences, Beijing 100190, China}
\affiliation{School of Physical Sciences, University of Chinese Academy of Sciences, Beijing 100049, China}

\author{
Jing Liu
}
\email{jing.liu@itp.ac.cn}
\affiliation{Institute of Theoretical Physics, Chinese Academy of Sciences, Beijing 100190, China}

\author{
Youjin Deng
}
\email{yjdeng@ustc.edu.cn}
\affiliation{Hefei National Laboratory for Physical Sciences at the Microscale and Department of Modern Physics, University of Science and Technology of China, Hefei 230026, China}
\affiliation{Hefei National Laboratory, University of Science and Technology of China, Hefei 230088, China}

\author{
Pan Zhang
}
\email{panzhang@itp.ac.cn}
\affiliation{Institute of Theoretical Physics, Chinese Academy of Sciences, Beijing 100190, China}

\date{\today}

\begin{abstract}
Efficient sampling of two-dimensional statistical physics systems remains a central challenge in computational statistical physics.
Traditional Markov chain Monte Carlo (MCMC) methods, including cluster algorithms, provide only partial solutions, as their efficiency collapses for large systems in the presence of frustration and quenched disorder.
The recently proposed Tensor Network Monte Carlo (TNMC) method offers a promising alternative, yet its original implementation suffers from inefficiencies due to the lack of scalable parallel sampling.

In this work, we introduce \textsc{BatchTNMC}, a GPU-optimized and parallelized implementation of TNMC tailored for large-scale simulations of two-dimensional spin glasses.
By leveraging batch processing and parallel sampling across multiple disorder realizations, our implementation achieves speedups of up to five orders of magnitude compared with the original serial scheme.
Benchmarking on two-dimensional spin glasses demonstrates dramatic gains in efficiency: for instance, {\bf{on a single GPU, \textsc{BatchTNMC} concurrently produces $\bf{1000}$ uncorrelated and unbiased samples across $\bf{1000}$ disorder realizations on $\bf{1024\times 1024}$ lattices in just 3.3 hours}}, with an acceptance probability of 37\%.
These results establish \textsc{BatchTNMC} as a scalable and powerful computational framework for the study of two-dimensional disordered spin glass systems.

\end{abstract}

\maketitle

\section{Introduction}

A central challenge in statistical physics lies in the accurate estimation of the partition function, the computation of thermodynamic observables, and the unbiased sampling of equilibrium configurations from the Boltzmann distribution.
The intrinsically high-dimensional nature of many-body systems makes these tasks computationally demanding, especially in frustrated and disordered systems such as spin glasses~\cite{sfedwards1975theory,binder1986spin}, where the rugged and complex energy landscape severely hinders conventional numerical approaches.
As spin glasses serve as paradigmatic models for understanding phase transitions and critical phenomena, the development of efficient and scalable computational approaches remains a pressing objective in the field of computational statistical physics.

Markov chain Monte Carlo (MCMC) algorithms~\cite{landau2014guide,newman1999monte} have long been indispensable tools for sampling from complex probability distributions, especially in the study of spin glasses.
Traditional schemes such as the Metropolis algorithm~\cite{metropolis1953equation}, which rely on local spin updates, suffer from critical slowing down near phase transitions, thereby limiting their efficiency.
More advanced methods, including parallel tempering~\cite{swendsen1986replica,e.marinari1992simulated,hukushima1996exchange}, alleviate these limitations to some extent but incur significant computational overhead, often restricting simulations to modest system sizes.
Cluster algorithms such as Swendsen-Wang~\cite{swendsen1987nonuniversal} and Wolff algorithm~\cite{wolff1989collective} succeed in mitigating critical slowing down for unfrustrated systems by introducing non-local cluster updates.
However, in frustrated systems their efficiency deteriorates sharply, as quenched randomness in the couplings suppresses the formation of meaningful clusters.
This fundamental limitation exposes a substantial gap in existing numerical methodologies for disordered spin glass systems.

Recent advances in machine learning, particularly in deep generative modeling, have introduced new computational paradigms into statistical physics~\cite{tomczak2022deep,wanglei2025computation}.
Variational autoregressive networks (VANs) have emerged as promising alternatives to conventional sampling techniques~\cite{wu2019solving}.
VANs employ autoregressive neural networks~\cite{germain2015made,oord2016pixel,oord2016conditional} as expressive variational ans\"atze, enabling efficient ancestral sampling~\cite{bishop2006pattern} without the need for Markov chains.
A salient advantage of VANs is their ability to generate statistically independent samples, which can be incorporated into importance sampling~\cite{nicoli2020asymptotically} or used as global spin updates and further refined by a Metropolis acceptance-rejection criterion, ensuing unbiased sampling from the Boltzmann distribution~\cite{mcnaughton2020boosting}.
Despite their promise, however, applications of generative neural networks have thus far remained largely limited to small-scale demonstrations, constrained by the considerable computational costs associated with training~\cite{liu2025efficient}.

Another recent line of progress involves hybrid approaches that incorporate tensor networks into Monte Carlo frameworks~\cite{frias-perez2023collective,chen2025tensor}.
Originally developed for quantum many-body simulations~\cite{white1992density,schollwock2011densitymatrix,orus2014practical,cirac2021matrix,orus2019tensor}, tensor networks have since found broad applications in statistical physics~\cite{levin2007tensor,liu2021tropical} and machine learning~\cite{cheng2019tree,cheng2021supervised,glasser2019expressive,han2018unsupervised,liu2023tensor,stoudenmire2016supervised}.
In particular, the partition function of the Ising model can be reformulated as a tensor network contraction problem.
Building on this observation, a pioneering study introduced a tensor-network-based sampling scheme in which ancestral sampling is performed through contractions encoding conditional probabilities~\cite{frias-perez2023collective}.
Unlike VANs, this approach dispenses with costly neural-network training, since contractions are deterministic with controllable numerical errors governed by bond dimensions and truncation thresholds.
The configurations generated from tensor network contractions can be utilized as global proposals within a Metropolis framework.
This Tensor Network Monte Carlo (TNMC) method was first validated on small systems as a proof of concept, and more recently demonstrated remarkable scalability: Ref.~\cite{chen2025tensor} applied TNMC to the two-dimensional random-bond Ising model~\cite{nishimori1981internal}, a paradigmatic and notoriously challenging spin glass, enabling simulations on lattices up to $1024 \times 1024$ and yielding precise estimates of multicritical points as well as bulk and surface exponents along the Nishimori line.

Despite these advances, the original TNMC implementation in Ref.~\cite{chen2025tensor} remains computationally suboptimal.
In spin glass simulations, quenched disorder averaging requires generating multiple samples for each disorder realization in order to construct the Markov chain.
The original scheme adopted a serial sampling strategy, producing one configuration at a time per realization. 
Although partial reuse of intermediate contractions was possible, this inherently sequential procedure led to inefficient utilization of modern computational resources.
Motivated by batch-processing techniques common in neural-network training, we propose a GPU-accelerated, parallel sampling framework for TNMC.
By introducing an additional computational dimension corresponding to disorder realizations, our method enables simultaneous sampling across many disorder instances, thus dramatically enhancing the overall efficiency.

In this work, we extend the TNMC framework of Ref.~\cite{chen2025tensor} and present \textsc{BatchTNMC}~\cite{batchtnmc}, a GPU-optimized implementation designed for large-scale two-dimensional spin glasses simulations.
Developed within the \textsc{PyTorch} ecosystem~\cite{paszke2019pytorch}, our implementation fully exploits the parallel processing capabilities of modern GPUs, achieving speedups of up to five orders of magnitude compared to the serial codes.
The rest of this paper is organized as follows:
Section~\ref{section:tnmc} outlines the theoretical foundations of TNMC; 
Section~\ref{section:code} describes the parallelization strategies and implementation details of \textsc{BatchTNMC}; 
Section~\ref{section:results} presents benchmarking results that highlight the substantial performance gains of our approach; 
and Section~\ref{section:conclusion} summarizes our findings and discusses future research directions.

\section{Principles of the Tensor Network Monte Carlo method}
\label{section:tnmc}

\subsection{Ancestral sampling from the Boltzmann distribution}

\begin{figure}[!tbp]
    \label{fig:tn}
    \centering
    \includegraphics[width=\textwidth]{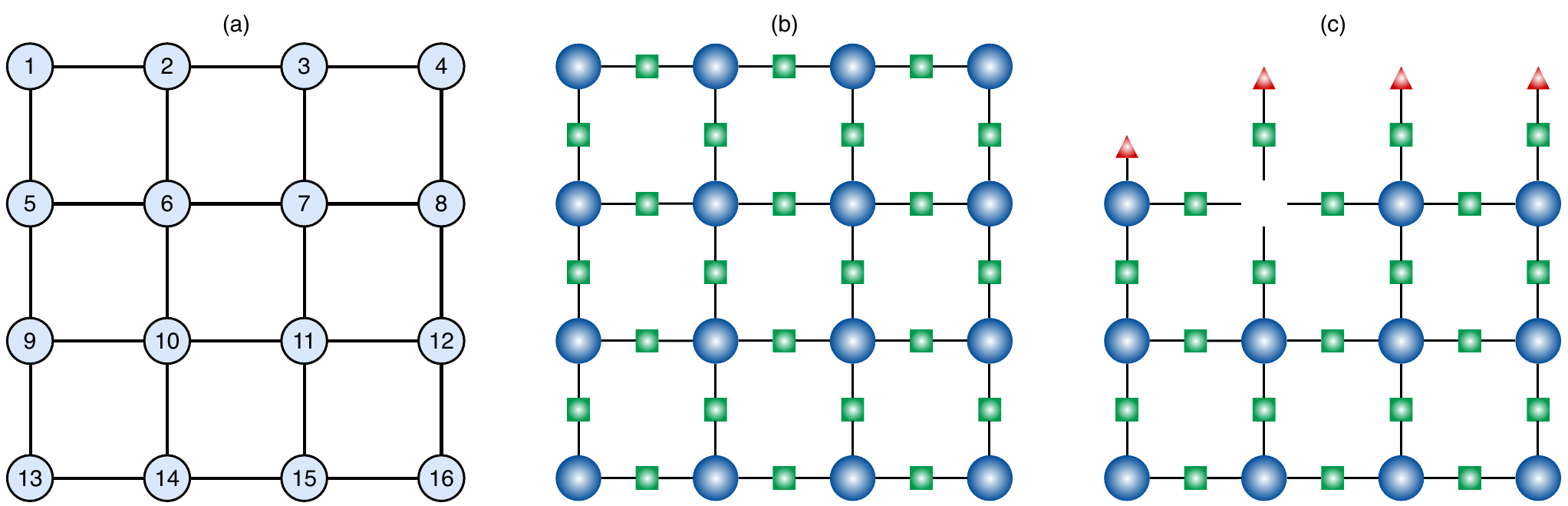}
    \caption{Tensor network representation of the two-dimensional Ising spin glass model.
    (a) Open square lattice of Ising spins, indexed in raster-scan order.
    (b) Partition function formulated as a tensor network contraction, where circles denote copy tensors and squares correspond to Boltzmann matrices.
    (c) Conditional partition function $f(s_6, \s_{<6})$ represented as a tensor network, with triangles indicating sampled spin variables.
    }
\end{figure}

To establish the theoretical foundation of the TNMC algorithm, we consider the Ising spin glass defined on an open square lattice of size $N = L \times L$.
The energy of a configuration $\s = \{s_1,\cdots,s_N\}$ is given by the Hamiltonian:
\begin{align}
    E(\s) = -\sum_{\langle i,j \rangle} J_{ij} s_i s_j,
\end{align}
where $s_i \in \{-1,+1\}$ denotes the spin at site $i$, $J_{ij}$ represents nearest-neighbor coupling, and $\langle i,j \rangle$ denotes the set of nearest-neighbor pairs.
The corresponding Boltzmann distribution reads
\begin{align}
    P(\s) = \frac{1}{Z} e^{-\beta E(\s)},
\end{align}
where $\beta$ is the inverse temperature and $Z = \sum_{\s} e^{-\beta E(\s)}$ is the partition function.
By applying the chain rule of probability, the joint distribution can be decomposed into a product of conditional probabilities,
\begin{align}
    P(\s) = P(s_1) \prod_{i=2}^{N} P(s_i | \s_{<i}),
\end{align}
where $\s_{<i} = \{s_1,\cdots,s_{i-1}\}$ and $\s_{>i} = \{s_{i+1}, \cdots, s_N\}$.
Throughout this work we adopt a raster-scan ordering on the two-dimensional lattice, see Fig.~\ref{fig:tn}(a).
This factorization is exact, and the conditional probability of spin $s_i$ is given by
\begin{align}
    P_i(s_i|\s_{<i}) = \frac{P(\s_{<i+1})}{P(\s_{<i})} = \frac{\sum_{\s_{>i}}e^{-\beta E(\s)}}{\sum_{\s_{>i-1}}e^{-\beta E(\s)}} = \frac{f(s_i,\s_{<i})}{\sum_{s_i}f(s_i,\s_{<i})},
\end{align}
where $f(s_i,\s_{<i}) = \sum_{\s_{>i}}e^{-\beta E(\s)}$ is the \textit{conditional partition function}.
The probability of fixing $s_i=+1$ given the preceding spins $\s_{<i}$ can be expressed compactly in terms of the sigmoid function $\sigma(x) = (1+e^{-x})^{-1}$:
\begin{align}
    P_i(s_i=+1|\s_{<i}) = \sigma\left(\log f(s_i=+1,\s_{<i}) - \log f(s_i=-1,\s_{<i})\right).
\end{align}
If the conditional partition functions can be evaluated, one may generate independent samples autoregressively via ancestral sampling~\cite{bishop2006pattern}: starting from $s_1$ from $P(s_1)$, then $s_2$ from $P(s_2 | s_1)$, and proceeding sequentially until the full configuration $\s$ is drawn.
This guarantees exact sampling from the Boltzmann distribution.

In practice, however, computing $P_i(s_i|\s_{<i})$ or $f(s_i,\s_{<i})$ is intractable for large $N$, as evaluating the sum over $\s_{>i}$ entails an exponentially large configuration space.
Indeed, exact sampling from the Boltzmann distribution belongs to the computational class \#-P, for which no polynomial-time algorithms are known.
In the following, we show how tensor-network methods provide efficient approximations to the conditional probabilities, thereby enabling scalable sampling from equilibrium distributions.

\subsection{Computing the conditional partition function via tensor network contraction}

Tensor networks provide a powerful formalism for representing and efficiently evaluating partition functions of statistical mechanical systems~\cite{levin2007tensor}.
For the Ising model, the partition function can be written as a sum over products of local Boltzmann factors,
\begin{align}
    Z = \sum_{\s} e^{-\beta E(\s)} = \sum_{\s} \prod_{\langle i,j \rangle} e^{\beta J_{ij} s_i s_j}.
\end{align}
This structure admits a natural tensor network representation, in which each nearest-neighbor interaction is encoded as a rank-2 tensor (matrix).
Concretely, for an interacting spin pair $(s_i, s_j)$, the Boltzmann factor is represented by the Boltzmann matrix
\begin{align}
    B = 
    \begin{pmatrix}
        e^{\beta J_{ij}} & e^{-\beta J_{ij}} \\
        e^{-\beta J_{ij}} & e^{\beta J_{ij}}
    \end{pmatrix},
\end{align}
with rows and columns corresponding to the states $s_i, s_j \in \{-1, +1\}$.
Summation over spin configurations corresponds to contracting shared indices between adjacent tensors.
Since each spin interacts with multiple neighbors, one must additionally enforce consistency across bonds connected to the same spin. This is achieved through a copy tensor $T$, which acts as a Kronecker delta ensuring
identical spin states on all attached indices.
For instance, a rank-4 copy tensor is defined as
\begin{align}
    T_{ijkl} = \begin{cases}
        1, & \text{if } i = j = k = l, \\
        0, & \text{otherwise.}
    \end{cases}
\end{align}
A schematic representation of this tensor network construction is illustrated in Fig.~\ref{fig:tn}(b), where circles denote copy tensors and squares denote the Boltzmann matrices.

The conditional partition function $f(s_i,\s_{<i})$ can be expressed in the same framework.
Fixing the spins in $\s_{<i}$ corresponds to clamping the associated indices in the tensor network to specific values.
The remaining spins $\s_{>i}$ are summed over, yielding a reduced contraction whose value is precisely $f(s_i, \s_{<i})$.
An illustrative example for $f(s_6, \s_{<6})$ is shown in Fig.~\ref{fig:tn}(c), where the triangles corresponding to $s_1,\cdots,s_5$ are fixed.

\begin{figure}[!tbp]
    \label{fig:bmps}
    \centering
    \includegraphics[width=\textwidth]{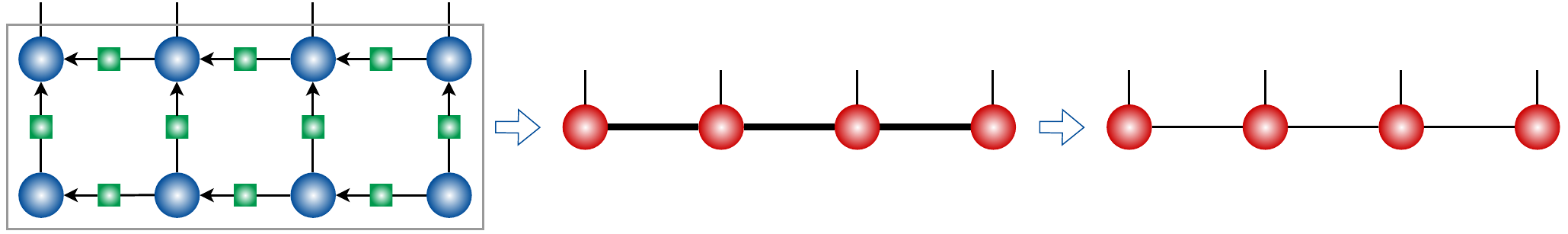}
    \caption{Multiplication of the Matrix Product Operator to the current Matrix Product State (MPS) yields a new MPS with doubled bond dimension. 
    The resulting MPS is subsequently truncated via singular value decomposition.}
\end{figure}

Exact contraction of two-dimensional tensor networks is intractable for large systems, scaling exponentially with lattice size $L$.
Practical simulations therefore rely on approximate contraction techniques. 
A widely adopted strategy is the boundary Matrix Product State (bMPS) method, which represents the boundary of the two-dimensional network as an MPS and iteratively contracts it with layers of tensors represented as Matrix Product Operators (MPOs), see Fig.~\ref{fig:bmps}.
Each contraction increases the bond dimension of the MPS, rapidly inflating computational cost. 
To control this growth, the MPS must be truncated after each step. 
This is accomplished using the singular value decomposition (SVD), which compresses the MPS by discarding subleading singular values while retaining the dominant entanglement structure~\cite{schollwock2011densitymatrix,orus2014practical}.

\subsection{Combining Metropolis scheme}

The approximate contraction method described above leads to an effective probability distribution of the autoregressive form
\begin{align}
    q(\s) = q(s_1) \prod_{i=2}^N q(s_i | \s_{<i}),
\end{align}
where each conditional probability $q(s_i | \s_{<i})$ is obtained from tensor network contractions~\cite{frias-perez2023collective,chen2025tensor}.
Although this structure enables efficient ancestral sampling, it introduces a systematic deviation from the exact Boltzmann distribution $P(\s)$.
To eliminate this bias, we employ a Metropolis–Hastings correction step~\cite{metropolis1953equation}.
Specifically, a candidate configuration $\s'$ is generated by ancestral sampling from the proposal distribution $q(\s)$.
The candidate is then accepted with probability
\begin{align}
    \label{eq:p_acc}
    P_{\text{acc}}(\s \rightarrow \s') = \min\left[1, \frac{q(\s) \times P(\s')}{q(\s') \times P(\s)}\right] = \min\left[1, \frac{q(\s) \times e^{-\beta E(\s')}}{q(\s') \times e^{-\beta E(\s)}}\right]
\end{align}
This procedure constitutes a global Markov chain update: each trial configuration $\s'$ is generated independently, in contrast to conventional local-update algorithms that alter only a small subset of spins.
The resulting Markov chain is ergodic and asymptotically converges to the exact Boltzmann distribution $P(\s)$, provided the support of $q(\s)$ includes all configurations with non-zero probability under $P(\s)$.

\section{Implementation of the \textsc{BatchTNMC}}
\label{section:code}

The \textsc{BatchTNMC} implementation is developed in \textsc{Python} with a modular and extensible architecture design to enhance its applicability across diverse two-dimensional Ising spin glass models.
The framework is organized into distinct computational components, including instance generation, cache initialization, and row-wise sampling.
This structured design enhances transparency and reproducibility while facilitating straightforward integration, modification, and extension for diverse applications in computational statistical physics.

\subsection{Generation of spin glass instances}
\label{subsection:spin_glass_instance}

The generation of different spin glass instances constitutes a critical preprocessing step in the \textsc{BatchTNMC} framework.
As a representative case, we consider the two-dimensional random-bond Ising model~\cite{nishimori1981internal}, implemented in the function \texttt{create\_rbim}.
The procedure begins with the initialization of a pseudorandom number generator (RNG) using a user-specified seed, thereby guaranteeing reproducibility of the generated disorder realizations.
A graph-based representation of the lattice is constructed via the \texttt{networkx.grid\_2d\_graph} routine, which systematically encodes the spatial connectivity among spins.
For each edge $(i, j)$ in the graph, the coupling $J_{ij}$ is independently sampled from a binary distribution, taking the value $-1$ with probability $p$ and $+1$ with probability $1-p$. Special cases include the pure ferromagnetic Ising model ($p=0$) and the Edwards–Anderson spin glass ($p=0.5$).

To enhance computational efficiency and enable direct integration with tensor-network operations, the graph representation is subsequently transformed into a three-dimensional PyTorch tensor $J_{\text{mat}}$ of shape $(L, L, 4)$ using the auxiliary function \texttt{graph\_to\_coupling} provided in \texttt{utils.py}.
Each tensor element $J_{\text{mat}}[i, j]$ encodes the set of four nearest-neighbor couplings (left, up, right, and down) associated with the spin at lattice site $(i, j)$.
For sites on open boundaries, absent neighbors are assigned a coupling value of zero.
Finally, the coupling tensor $J_{\text{mat}}$ is saved to disk via PyTorch's \texttt{torch.save} utility, ensuring data reusability across computational sessions.

\begin{lstlisting}[language=python]
def create_rbim(L, p, seed=1):
    rng = np.random.default_rng(seed)
    graph = nx.grid_2d_graph(L, L)
    for u, v in graph.edges():
        graph[u][v]["weight"] = rng.choice([-1.0, 1.0], p=[p, 1 - p])
    coupling_mat = graph_to_coupling(graph, L)
    filename = f"./instances/L{L}_p{p:.6f}_seed{seed}.pt"
    os.makedirs(os.path.dirname(filename), exist_ok=True)
    torch.save(coupling_mat.float(), filename)
\end{lstlisting}

\subsection{Cache Initialization via boundary MPS}
\label{subsection:cache}
We now describe the implementation of the cache initialization procedure based on the bMPS algorithm.
Starting from the bottom boundary of the lattice, the MPS is iteratively constructed and compressed in order to control the growth of the bond dimension. The resulting MPS serves as a cache that encodes conditional partition functions, which are required for row-by-row spin configuration sampling.

In our implementation, the MPS is represented as a Python list of length $L$, with each entry corresponding to a three-dimensional tensor with dimensions $[(1, 2, \chi), (\chi, 2, \chi), \dots, (\chi, 2, 1)]$. 
Here, $\chi$ denotes the maximum bond dimension, which directly determines both the representational accuracy and the computational cost.
The corresponding MPO is similarly represented as a list of four-dimensional tensors with dimensions $[(1, 2, 2, 2), (2, 2, 2, 2), \dots, (2, 2, 1, 2)]$.

The initialization begins with the construction of the bottom-row MPS. 
Boltzmann matrices are generated using the auxiliary function \texttt{set\_ising\_bmat} and subsequently combined with copy tensors to form the initial MPS.
The algorithm then proceeds iteratively from the penultimate row to the top of the lattice.
For each row, the corresponding MPO is first constructed and applied to the current MPS using the \texttt{multiply} function, which performs the contraction between an MPO and an MPS~\cite{schollwock2011densitymatrix,orus2014practical}.
This operation leads to a doubling of the bond dimension, which is subsequently truncated to the prescribed maximum $\chi$ using the \texttt{compress} function based on SVD. 
Repeating this procedure across all rows produces a complete cache of MPS representations for the lattice.

\begin{lstlisting}[language=python]
def create_cache_ising(L, chi, beta, p, seed):
    # load Hamiltonian
    ham_path = f"./instances/L{L}_p{p:.6f}_seed{seed}.pt"
    J_mat = torch.load(ham_path, weights_only=True)  # (L, L, 4)

    # define copy tensor for constructing the MPS and MPO
    I2, I3, I4 = torch.eye(2), torch.zeros((2, 2, 2)), torch.zeros((2, 2, 2, 2))
    for i in range(2): I3[i, i, i] = 1; I4[i, i, i, i] = 1

    mps = []
    B_mat = set_ising_bmat(beta, J_mat[-1, 0, 2])
    mps.append((I2 @ B_mat).unsqueeze(0))  # (1, 2, D)
    for col in range(1, L - 1):
        B_mat = set_ising_bmat(beta, J_mat[-1, col, 2]); mps.append((I3 @ B_mat))  # (D, 2, D)
    mps.append(I2.unsqueeze(-1))  # (D, 2, 1)
    cache_path = f"./cache/L{L}_chi{chi}_beta{beta:.6f}_p{p:.6f}_seed{seed}/row{L-1}.pt"
    torch.save([tensor.clone() for tensor in mps], cache_path)

    for row in tqdm.tqdm(range(L - 2, -1, -1), desc="Building MPS cache"):
        # create the MPO for current row
        mpo = []
        B_mat_d = set_ising_bmat(beta, J_mat[row, 0, 3])
        B_mat_r = set_ising_bmat(beta, J_mat[row, 0, 2])
        mpo.append((torch.einsum("ijk,jl,km->ilm", I3, B_mat_r, B_mat_d).unsqueeze(0)))  # (1, 2, 2, 2)
        for col in range(1, L - 1):
            B_mat_d = set_ising_bmat(beta, J_mat[row, col, 3])
            B_mat_r = set_ising_bmat(beta, J_mat[row, col, 2])
            mpo.append(torch.einsum("ijkl,km,ln->ijmn", I4, B_mat_r, B_mat_d))  # (2, 2, 2, 2)
        B_mat_d = set_ising_bmat(beta, J_mat[row, L - 1, 3])
        mpo.append((torch.einsum("ijk,kl->ijl", I3, B_mat_d).unsqueeze(2)))  # (2, 2, 1, 2)
        # MPO x MPS
        mps = multiply(mpo, mps)  
        _, mps = compress(mps, chi=chi)  # compress the new MPS
        cache_path = f"./cache/L{L}_chi{chi}_beta{beta:.6f}_p{p:.6f}_seed{seed}/row{row}.pt"
        torch.save([tensor.clone() for tensor in mps], cache_path)
\end{lstlisting}

\subsection{Tensor network contraction for conditional probabilities}

We now describe the implementation of the tensor network contraction scheme used to compute the conditional probabilities for sampling.
This component integrates both the disorder instances generated in Section~\ref{subsection:spin_glass_instance} and the MPS cache initialized in Section~\ref{subsection:cache}.
The functionality is encapsulated within the Python class called \texttt{SamplerIsing}, which first loads the spin-glass coupling matrix from disk and transfers it to the designated computational device.

\begin{lstlisting}[language=python]
class SamplerIsing:
    def __init__(self, L, chi, beta, p, seed, device="cuda:0"):
        self.L = L; self.chi=chi; self.beta=beta; self.p=p; self.seed=seed; self.device=device

        ham_path = f"./instances/L{L}_p{p:.6f}_seed{seed}.pt"
        self.J_mat = torch.load(ham_path, weights_only=True, map_location=self.device)

    def sample(self, bs=1000):
        cfgs = torch.zeros((bs, self.L, self.L), dtype=torch.booL, device=self.device)
        log_probs = torch.zeros(bs, dtype=torch.float, device=self.device)
        for row in tqdm(range(self.L), desc="Sampling row"):
            cache_path =
        f"./cache/L{self.L}_chi{self.chi}_beta{self.beta:.6f}_p{self.p:.6f}_seed{self.seed}/row{row}.pt"
            mps = torch.load(cache_path, map_location=self.device, weights_only=True)
            cfg_up = cfgs[:, row - 1, :]; J_mat_row = self.J_mat[row, :, :]  # (bs, L) and (L, 4)
            row_cfgs, row_log_probs = self._sample_row(bs, mps, cfg_up, J_mat_row)
            cfgs[:, row, :] = row_cfgs
            log_probs += row_log_probs
        return cfgs, log_probs
    
    def _sample_row(self, bs, mps, cfg_up, J_mat_row):
        row_cfgs = torch.zeros((bs, self.L), dtype=torch.bool, device=self.device)
        row_log_probs = torch.zeros((bs, self.L), dtype=torch.float, device=self.device)
        # build right environment given the configurations of the upper row and Boltzmann weights
        right_envs = {}  # right_envs[j] is the right env tensor for sampling the j-th column, (bs, D)
        right_envs[self.L - 1] = torch.ones(bs, 1, dtype=torch.float, device=self.device)
        for col in range(self.L - 2, -1, -1):
            B_mat = set_ising_bmat(self.beta, J_mat_row[col + 1, 1], dtype=torch.float, device=self.device)
            tmp = torch.einsum("ijk,lj->ilk", mps[col + 1], B_mat)
            tmp = tmp[:, cfg_up[:, col + 1].long(), :]  # (D_l, bs, D_r)
            right_envs[col] = torch.einsum("ibj,bj->bi", tmp, right_envs[col + 1])  # update right env
            right_envs[col] /= torch.norm(right_envs[col], dim=1, keepdim=True)
        # sampling from left to right
        left_env = torch.ones((bs, 1), dtype=torch.float, device=self.device)
        for col in range(self.L):
            rho = contract("bi,ijk,bk->bj", left_env, mps[col], right_envs[col]).abs()  # (bs, 2)
            up_field = 2 * self.beta * J_mat_row[col, 1] * (2 * cfg_up[:, col] - 1)
            probs = torch.sigmoid(torch.log(rho[:, 1] / rho[:, 0]) + up_field)
            dist = Bernoulli(probs=probs)
            row_cfgs[:, col] = dist.sample()
            row_log_probs[:, col] = dist.log_prob(row_cfgs[:, col])
            left_env = torch.einsum("bi,ibj->bj", left_env, mps[col][:, row_cfgs[:, col].long(), :])
            left_env /= torch.norm(left_env, dim=1, keepdim=True)
        return row_cfgs, row_log_probs.sum(dim=1)
\end{lstlisting}

The central sampling procedure is realized through the methods \texttt{sample()} and \texttt{\_sample\_row()}.
The interface accepts a batch size parameter \texttt{bs}, allowing for the parallel generation of multiple statistically independent configurations corresponding to a single disorder realization. For each row, the corresponding MPS cache is retrieved, and conditional probabilities are computed using the previously sampled row (\texttt{cfg\_up}) together with the row-specific coupling matrix (\texttt{J\_mat\_row}). The outputs are two tensors: \texttt{row\_cfgs}, of shape \texttt{(bs, L)}, which stores the sampled spin configurations of the current row, and \texttt{row\_log\_probs}, which records the associated log-probabilities.

Prior to sampling, right environments \texttt{right\_envs} are precomputed for all columns of the current row.
The right environment of column $j$, denoted as \texttt{right\_envs[j]}, corresponds to the contracted tensor network to the right of site $j$.
These environments are constructed by contracting the appropriate tensor from the MPS cache with the Boltzmann matrix (\texttt{B\_mat}) and the upper-row configuration (\texttt{cfg\_up}).
Once the right environments are available, spins are sampled sequentially from left to right across the row according to the following procedure:
\begin{enumerate}
    \item Probability computation: The left environment (\texttt{left\_env}) is contracted with the boundary MPS tensor of the current column and the corresponding right environment \texttt{right\_envs[j]} to evaluate the probability for the spin at column $j$.
    \item Configuration sampling: A Bernoulli distribution is constructed from this probability, and independent samples are drawn to determine the spin configuration at column $j$.
    \item Log-probability storage: The log-probability of the sampled configuration is computed and stored for subsequent use.
    \item Environment update: The left environment is updated by contracting the current MPS tensor with the sampled spin value. The updated tensor is normalized to ensure numerical stability during the propagation.
\end{enumerate}

\subsection{Handling multi-disorder systems}

In the preceding sections, we have demonstrated the core implementation of the \textsc{BatchTNMC} using a single disorder realization as a pedagogical example.
For a given realization, the generation of multiple statistically independent spin configurations can already be parallelized efficiently. 
Extending this framework to multiple disorder realizations, however, poses an additional challenge. 
A straightforward approach would be to process disorder instances sequentially in iterative loops, which is computationally inefficient.

To overcome this limitation, \textsc{BatchTNMC} adopts a fully batched strategy in which an auxiliary dimension is introduced to index disorder realizations. 
This design enables the simultaneous and parallel processing of multiple disorder instances, and when combined with GPU acceleration, results in substantial performance gains.
Concretely, the coupling matrix $J_{\mathrm{mat}}$ is promoted to a four-dimensional tensor of shape $(n_{\mathrm{dis}}, L, L, 4)$, where the leading dimension $n_{\mathrm{dis}}$ enumerates distinct disorder configurations. 
The MPS cache retains its list-based structure of length $L$, but each tensor is augmented with an additional leading dimension, taking the form $[(n_{\mathrm{dis}}, 1, 2, \chi), (n_{\mathrm{dis}}, \chi, 2, \chi), \dots, (n_{\mathrm{dis}}, \chi, 2, 1)]$.
This consistent incorporation of the $n_{\mathrm{dis}}$ dimension propagates naturally throughout the workflow, including in the ancestral sampling stage, where conditional probabilities are evaluated simultaneously across all disorder realizations. 

\subsection{Computational complexity of the TNMC}

Before presenting the numerical results, it is instructive to analyze the computational complexity of the TNMC algorithm.
Let the physical dimension be denoted by $d$ (with $d=2$ for the Ising model).
In the bMPS-based cache initialization, the dominant computational costs originate from two sources: (i) the multiplication of the MPO with the MPS, which scales as $\mathcal{O}(L^2 \chi^2 d^4)$, and (ii) the SVD procedure, which scales as $\mathcal{O}(L^2 \chi^3)$.
Once constructed, the cache can be reused for sampling without further recomputation.
The computational cost of sampling a single configuration consists of two main contributions: (i) the construction of right environments, scaling as $\mathcal{O}(\chi^2)$, and (ii) the subsequent sampling step, which scales as $\mathcal{O}(\chi^2 d)$. 

In practical applications involving large system sizes or extensive disorder realizations, memory constraints may arise when storing the complete MPS cache or sampled configurations on GPU devices. 
To mitigate this issue, one can incorporate a GPU memory-efficient streaming protocol. 
Rather than retaining the entire MPS cache in device memory, tensors associated with individual rows are stored independently. 
During ancestral sampling, only the tensors required at a given stage are dynamically loaded into GPU memory row by row or site by site.
This strategy substantially reduces memory overhead while preserving computational efficiency. 

\section{Numerical results}
\label{section:results}

To assess the performance of \textsc{BatchTNMC}, we apply it to the two-dimensional EA model, a canonical testbed for spin glasses.
Within the TNMC framework, we evaluate the Metropolis acceptance probability $P_{\text{acc}}$ [Eq.~\ref{eq:p_acc}], which governs whether a proposed transition between spin configurations is accepted in the global MCMC procedure.
A high acceptance probability implies that nearly all proposed configurations are accepted, thereby minimizing wasted computational effort.
In the limiting case $P_{\text{acc}}=1$, the proposal distribution coincides exactly with the Boltzmann distribution, ensuring unbiased estimates of thermodynamic observables.

\begin{figure}[!tbp]
    \label{fig:acceptance}
    \centering
    \includegraphics[width=\textwidth]{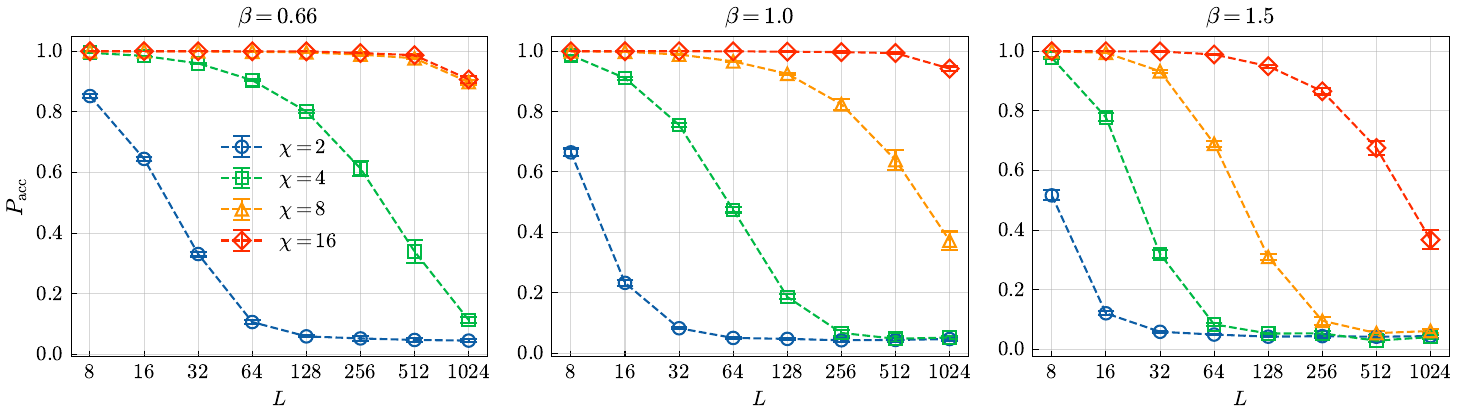}
    \caption{Acceptance probability $P_{\text{acc}}$ of the TNMC algorithm for the two-dimensional Edwards-Anderson model, evaluated for different system sizes $L$ and bond dimensions $\chi$ at inverse temperatures $\beta=0.66, 1.0$, and $1.5$.}
\end{figure}

We investigate system sizes ranging from $L=8$ to $1024$ at fixed inverse temperatures $\beta=0.66, 1.0$, and $1.5$.
As shown in Fig.~\ref{fig:acceptance}, the acceptance probability decays with increasing system size.
Nevertheless, practical acceptance probabilities are preserved even for moderate bond dimensions. 
For instance, at $L=1024$ and $\beta=1.0$, a bond dimension of $\chi=8$ yields $P_{\text{acc}} \approx 0.37$, while for $L=1024$ and $\beta=1.5$, increasing to $\chi=16$ maintains a comparable probability. 
As a concrete demonstration, for $\beta=1.0$ and $\chi=8$, \textsc{BatchTNMC} generates $10^{3}$ uncorrelated  and unbiased samples across $10^{3}$ disorder realizations at system size $1024 \times 1024$ within 3.3 hours on a single GPU, requiring approximately 52 GB of GPU memory.

\begin{figure}[!tbp]
    \label{fig:speedup}
    \centering
    \includegraphics[width=0.75\textwidth]{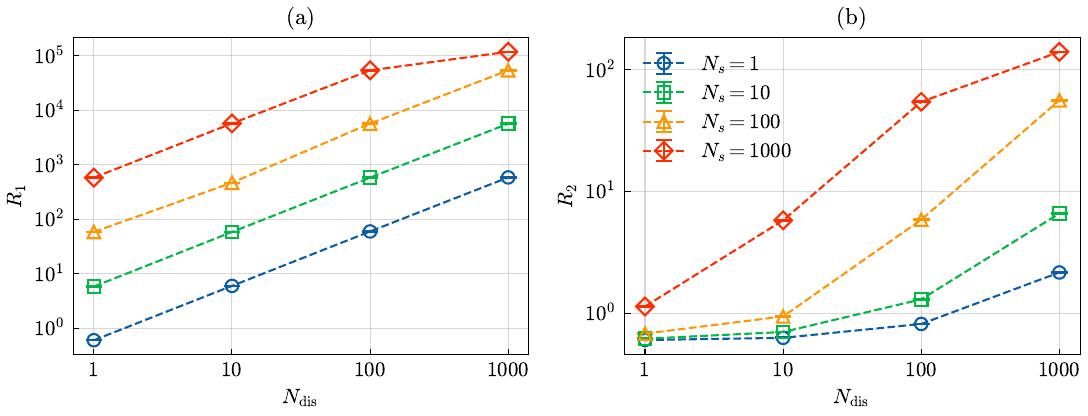}
    \caption{
    Speedup relative to (a) the baseline serial implementation reported in Ref.~\cite{chen2025tensor} and (b) the CPU-based counterpart, evaluated for system size $L = 1024$ and bond dimension $\chi = 8$ as a function of batch size $N_s$ and the number of disorder realizations $N_{\text{dis}}$.}
\end{figure}

We further quantify the computational efficiency of \textsc{BatchTNMC} by defining two speedup ratios.
The first compares GPU-accelerated sampling to the serial CPU implementation of Ref.~\cite{chen2025tensor}:
\begin{align}
    R_1 = \frac{N_{\text{s}} \times N_{\text{dis}} \times T_{\text{serial}}(L, \chi)}{T_{\text{GPU}}(L, \chi, N_{\text{s}}, N_{\text{dis}})},
\end{align}
where $T_{\text{serial}}(L, \chi)$ is the runtime of the serial implementation that processes one sample per disorder realization sequentially, and $T_{\text{GPU}}(L, \chi, N_{\text{s}}, N_{\text{dis}})$ is the runtime of \textsc{BatchTNMC} on GPU for $N_{\text{s}}$ samples across $N_{\text{dis}}$ realizations.
The second speedup ratio measures GPU acceleration relative to the CPU implementation of \textsc{BatchTNMC}:
\begin{align}
    R_2 = \frac{T_{\text{CPU}}(L, \chi, N_{\text{s}}, N_{\text{dis}})}{T_{\text{GPU}}(L, \chi, N_{\text{s}}, N_{\text{dis}})},
\end{align}
where $T_{\text{CPU}}(L, \chi, N_{\text{s}}, N_{\text{dis}})$ denotes the runtime on CPU.
The results, summarized in Fig.~\ref{fig:speedup} for $L=1024$, demonstrate that both $R_{1}$ and $R_{2}$ increase with $N_{\text{s}}$ and $N_{\text{dis}}$, reaching values up to $10^{5}$ relative to the serial implementation. 
Furthermore, the GPU-accelerated \textsc{BatchTNMC} achieves a performance acceleration exceeding two orders of magnitude compared to its CPU counterpart.

\section{Conclusion}
\label{section:conclusion}

In this work, we have presented \textsc{BatchTNMC}, an optimized and parallelized implementation of the tensor network Monte Carlo (TNMC) algorithm, specifically tailored for large-scale, GPU-accelerated simulations of two-dimensional spin glasses.
By systematically exploiting batch processing and parallel sampling across multiple disorder realizations, \textsc{BatchTNMC} achieves substantial computational gains. 
Benchmarking results demonstrate that, relative to the original serial implementation, \textsc{BatchTNMC} attains speedups of up to five orders of magnitude, while its GPU-accelerated variant surpasses the CPU implementation by approximately a factor of $10^2$.
These advances enable simulations of unprecedented scale, extending TNMC to system sizes and disorder ensembles that would be computationally prohibitive with conventional methods.
Beyond performance optimization, the framework provides a modular and extensible foundation for investigating disordered spin systems. 
In future developments, we plan to generalize the package to encompass a broader class of two-dimensional models, such as the $q$-state Potts, as well as to extend the methodology to three-dimensional spin glasses, where sampling challenges are even more pronounced. 
Such extensions will further broaden the applicability of TNMC in computational statistical physics.

Our PyTorch implementation of \textsc{BatchTNMC} and tutorial notebooks on concurrently generating $1000$ unbiased samples across $1000$ disorder realizations on $1024\times 1024$ lattices is publicly available in Ref.~\cite{batchtnmc}.

\begin{acknowledgments}
This work is supported by Projects 12325501, 12247104, 12275263, and 12405047 of the National Natural Science Foundation of China, and the Innovation Program for Quantum Science and Technology (under Grant No. 2021ZD0301900). YD thanks the support from the Natural Science Foundation of Fujian Province of China (under Grant No. 2023J02032).
\end{acknowledgments}

\bibliographystyle{iopart-num.bst}
\bibliography{main}

\providecommand{\newblock}{}
\begin{thebibliography}{10}
\expandafter\ifx\csname url\endcsname\relax
  \def\url#1{{\tt #1}}\fi
\expandafter\ifx\csname urlprefix\endcsname\relax\def\urlprefix{URL }\fi
\providecommand{\eprint}[2][]{\url{#2}}

\bibitem{sfedwards1975theory}
{S F Edwards} and {P W Anderson} 1975 {\em Journal of Physics F: Metal Physics\/} {\bf 5} 965 ISSN 0305-4608 \urlprefix\url{https://dx.doi.org/10.1088/0305-4608/5/5/017}

\bibitem{binder1986spin}
Binder K and Young A~P 1986 {\em Reviews of Modern Physics\/} {\bf 58} 801--976 \urlprefix\url{https://link.aps.org/doi/10.1103/RevModPhys.58.801}

\bibitem{landau2014guide}
Landau D~P and Binder K 2014 {\em A Guide to Monte Carlo Simulations in Statistical Physics\/} 4th ed (Cambridge: Cambridge University Press) \urlprefix\url{https://www.cambridge.org/core/product/2522172663AF92943C625056C14F6055}

\bibitem{newman1999monte}
Newman M and Barkema G 1999 {\em Monte Carlo Methods in Statistical Physics\/} (Clarendon Press) ISBN 978-0-19-851796-2 \urlprefix\url{https://books.google.co.jp/books?id=KKL2nQEACAAJ}

\bibitem{metropolis1953equation}
Metropolis N, Rosenbluth A~W, Rosenbluth M~N, Teller A~H and Teller E 1953 {\em The Journal of Chemical Physics\/} {\bf 21} 1087--1092 ISSN 0021-9606 \urlprefix\url{https://doi.org/10.1063/1.1699114}

\bibitem{swendsen1986replica}
Swendsen R~H and Wang J~S 1986 {\em Physical Review Letters\/} {\bf 57} 2607--2609 \urlprefix\url{https://link.aps.org/doi/10.1103/PhysRevLett.57.2607}

\bibitem{e.marinari1992simulated}
{E Marinari} and {G Parisi} 1992 {\em Europhysics Letters\/} {\bf 19} 451 ISSN 0295-5075 \urlprefix\url{https://dx.doi.org/10.1209/0295-5075/19/6/002}

\bibitem{hukushima1996exchange}
Hukushima K and Nemoto K 1996 {\em Journal of the Physical Society of Japan\/} {\bf 65} 1604--1608 ISSN 0031-9015 \urlprefix\url{https://doi.org/10.1143/JPSJ.65.1604}

\bibitem{swendsen1987nonuniversal}
Swendsen R~H and Wang J~S 1987 {\em Physical Review Letters\/} {\bf 58} 86--88 \urlprefix\url{https://link.aps.org/doi/10.1103/PhysRevLett.58.86}

\bibitem{wolff1989collective}
Wolff U 1989 {\em Physical Review Letters\/} {\bf 62} 361--364 \urlprefix\url{https://link.aps.org/doi/10.1103/PhysRevLett.62.361}

\bibitem{tomczak2022deep}
Tomczak J~M 2022 {\em Deep Generative Modeling\/} (Cham: Springer International Publishing) ISBN 978-3-030-93157-5 978-3-030-93158-2 \urlprefix\url{https://link.springer.com/10.1007/978-3-030-93158-2}

\bibitem{wanglei2025computation}
Wang L and Zhang P 2025 {\em PHYSICS\/} {\bf 54} 10--18 ISSN 0379-4148 \urlprefix\url{https://wuli.iphy.ac.cn/en/article/doi/10.7693/wl20250102}

\bibitem{wu2019solving}
Wu D, Wang L and Zhang P 2019 {\em Physical Review Letters\/} {\bf 122} 080602 ISSN 0031-9007, 1079-7114 \urlprefix\url{https://link.aps.org/doi/10.1103/PhysRevLett.122.080602}

\bibitem{germain2015made}
Germain M, Gregor K, Murray I and Larochelle H 2015 Made: Masked autoencoder for distribution estimation {\em Proceedings of the 32nd International Conference on Machine Learning\/} ({\em Proceedings of Machine Learning Research\/} vol~37) ed Bach F and Blei D (Lille, France: PMLR) pp 881--889 \urlprefix\url{https://proceedings.mlr.press/v37/germain15.html}

\bibitem{oord2016pixel}
van~den Oord A, Kalchbrenner N and Kavukcuoglu K 2016 Pixel recurrent neural networks {\em Proceedings of the 33nd International Conference on Machine Learning, ICML 2016, New York City, NY, USA, June 19-24, 2016\/} ({\em JMLR Workshop and Conference Proceedings\/} vol~48) ed Balcan M~F and Weinberger K~Q (JMLR.org) pp 1747--1756 \urlprefix\url{http://proceedings.mlr.press/v48/oord16.html}

\bibitem{oord2016conditional}
van~den Oord A, Kalchbrenner N, Espeholt L, Kavukcuoglu K, Vinyals O and Graves A 2016 Conditional image generation with pixelcnn decoders {\em Advances in Neural Information Processing Systems 29: Annual Conference on Neural Information Processing Systems 2016, December 5-10, 2016, Barcelona, Spain\/} ed Lee D~D, Sugiyama M, von Luxburg U, Guyon I and Garnett R pp 4790--4798 \urlprefix\url{https://proceedings.neurips.cc/paper/2016/hash/b1301141feffabac455e1f90a7de2054-Abstract.html}

\bibitem{bishop2006pattern}
Bishop C~M 2006 {\em Pattern Recognition and Machine Learning\/} (Springer New York, NY)

\bibitem{nicoli2020asymptotically}
Nicoli K~A, Nakajima S, Strodthoff N, Samek W, M{\"u}ller K~R and Kessel P 2020 {\em Physical Review E\/} {\bf 101} 023304 ISSN 2470-0045, 2470-0053 \urlprefix\url{https://link.aps.org/doi/10.1103/PhysRevE.101.023304}

\bibitem{mcnaughton2020boosting}
McNaughton B, Milo{\v s}evi{\'c} M~V, Perali A and Pilati S 2020 {\em Physical Review E\/} {\bf 101} 053312 ISSN 2470-0045, 2470-0053 \urlprefix\url{https://link.aps.org/doi/10.1103/PhysRevE.101.053312}

\bibitem{liu2025efficient}
Liu J, Tang Y and Zhang P 2025 {\em Physical Review E\/} {\bf 111} 025304 \urlprefix\url{https://link.aps.org/doi/10.1103/PhysRevE.111.025304}

\bibitem{frias-perez2023collective}
{Fr{\'i}as-P{\'e}rez} M, Mari{\"e}n M, Garc{\'i}a D~P, Ba{\~n}uls M~C and Iblisdir S 2023 {\em SciPost Physics\/} {\bf 14} 123 \urlprefix\url{https://scipost.org/10.21468/SciPostPhys.14.5.123}

\bibitem{chen2025tensor}
Chen T, Guo E, Zhang W, Zhang P and Deng Y 2025 {\em Physical Review B\/} {\bf 111} 094201 \urlprefix\url{https://link.aps.org/doi/10.1103/PhysRevB.111.094201}

\bibitem{white1992density}
White S~R 1992 {\em Physical Review Letters\/} {\bf 69} 2863--2866 \urlprefix\url{https://link.aps.org/doi/10.1103/PhysRevLett.69.2863}

\bibitem{schollwock2011densitymatrix}
Schollw{\"o}ck U 2011 {\em Annals of Physics\/} {\bf 326} 96--192 ISSN 00034916 \urlprefix\url{https://linkinghub.elsevier.com/retrieve/pii/S0003491610001752}

\bibitem{orus2014practical}
Or{\'u}s R 2014 {\em Annals of Physics\/} {\bf 349} 117--158 ISSN 00034916 \urlprefix\url{https://linkinghub.elsevier.com/retrieve/pii/S0003491614001596}

\bibitem{cirac2021matrix}
Cirac J~I, {P{\'e}rez-Garc{\'i}a} D, Schuch N and Verstraete F 2021 {\em Reviews of Modern Physics\/} {\bf 93} 045003 ISSN 0034-6861, 1539-0756 \urlprefix\url{https://link.aps.org/doi/10.1103/RevModPhys.93.045003}

\bibitem{orus2019tensor}
Or{\'u}s R 2019 {\em Nature Reviews Physics\/} {\bf 1} 538--550 ISSN 2522-5820 \urlprefix\url{http://www.nature.com/articles/s42254-019-0086-7}

\bibitem{levin2007tensor}
Levin M and Nave C~P 2007 {\em Physical Review Letters\/} {\bf 99} 120601 \urlprefix\url{https://link.aps.org/doi/10.1103/PhysRevLett.99.120601}

\bibitem{liu2021tropical}
Liu J~G, Wang L and Zhang P 2021 {\em Physical Review Letters\/} {\bf 126} 090506 ISSN 0031-9007, 1079-7114 \urlprefix\url{https://link.aps.org/doi/10.1103/PhysRevLett.126.090506}

\bibitem{cheng2019tree}
Cheng S, Wang L, Xiang T and Zhang P 2019 {\em Physical Review B\/} {\bf 99} 155131 ISSN 2469-9950, 2469-9969 \urlprefix\url{https://link.aps.org/doi/10.1103/PhysRevB.99.155131}

\bibitem{cheng2021supervised}
Cheng S, Wang L and Zhang P 2021 {\em Physical Review B\/} {\bf 103} 125117

\bibitem{glasser2019expressive}
Glasser I, Sweke R, Pancotti N, Eisert J and Cirac I 2019 Expressive power of tensor-network factorizations for probabilistic modeling {\em Advances in Neural Information Processing Systems\/} vol~32 ed Wallach H, Larochelle H, Beygelzimer A, {dAlch{\'e}-Buc} F, Fox E and Garnett R (Curran Associates, Inc.) \urlprefix\url{https://proceedings.neurips.cc/paper_files/paper/2019/file/b86e8d03fe992d1b0e19656875ee557c-Paper.pdf}

\bibitem{han2018unsupervised}
Han Z~Y, Wang J, Fan H, Wang L and Zhang P 2018 {\em Physical Review X\/} {\bf 8} 031012 ISSN 2160-3308 \urlprefix\url{https://link.aps.org/doi/10.1103/PhysRevX.8.031012}

\bibitem{liu2023tensor}
Liu J, Li S, Zhang J and Zhang P 2023 {\em Physical Review E\/} {\bf 107} L012103 \urlprefix\url{https://link.aps.org/doi/10.1103/PhysRevE.107.L012103}

\bibitem{stoudenmire2016supervised}
Stoudenmire E and Schwab D~J 2016 Supervised learning with tensor networks {\em Advances in Neural Information Processing Systems\/} vol~29 ed Lee D, Sugiyama M, Luxburg U, Guyon I and Garnett R (Curran Associates, Inc.) \urlprefix\url{https://proceedings.neurips.cc/paper_files/paper/2016/file/5314b9674c86e3f9d1ba25ef9bb32895-Paper.pdf}

\bibitem{nishimori1981internal}
Nishimori H 1981 {\em Progress of Theoretical Physics\/} {\bf 66} 1169--1181 ISSN 0033-068X \urlprefix\url{https://doi.org/10.1143/PTP.66.1169}

\bibitem{batchtnmc}
\url{https://github.com/bnuliujing/BatchTNMC}

\bibitem{paszke2019pytorch}
Paszke A, Gross S, Massa F, Lerer A, Bradbury J, Chanan G, Killeen T, Lin Z, Gimelshein N, Antiga L, Desmaison A, Kopf A, Yang E, DeVito Z, Raison M, Tejani A, Chilamkurthy S, Steiner B, Fang L, Bai J and Chintala S 2019 Pytorch: An imperative style, high-performance deep learning library {\em Advances in Neural Information Processing Systems\/} vol~32 ed Wallach H, Larochelle H, Beygelzimer A, {dAlch{\'e}-Buc} F, Fox E and Garnett R (Curran Associates, Inc.) \urlprefix\url{https://proceedings.neurips.cc/paper_files/paper/2019/file/bdbca288fee7f92f2bfa9f7012727740-Paper.pdf}

\end{thebibliography}

\clearpage
\newpage

\end{document}